\documentclass[12pt]{iopart}

\usepackage{graphicx}
\usepackage{amssymb}
\usepackage{lineno}
\usepackage{mathrsfs}
\usepackage{siunitx}
\usepackage[latin1]{inputenc}
\usepackage{bbm}
\usepackage{graphicx}
\usepackage{graphics}
\usepackage{epsfig}
\usepackage{epstopdf}
\usepackage{xcolor}
\usepackage{mathrsfs}
\usepackage{float}
\usepackage{multirow}
\usepackage{ulem}

\begin{document}

%%%%% FRONTMATTER

\title[]{Optimum heat treatment to enhance the weak-link response of Y123 nanowires prepared by Solution Blow Spinning}

\author{Ana M. Caffer$^1$, Davi A. D. Chaves$^2$, Alexsander L. Pessoa$^1$, Claudio L. Carvalho$^1$, Wilson A. Ortiz$^2$, Rafael Zadorosny$^1$, and Maycon Motta$^2$}

\address{$^{1}$Departamento de F\' isica e Qu\' imica, Universidade Estadual Paulista (UNESP), Faculdade de Engenharia, Caixa Postal 31, 15385-000, Ilha Solteira, SP, Brazil

$^{2}$Departamento de F\'isica, Universidade Federal de S\~ao Carlos (UFSCar), S\~ao Carlos, SP, Brazil}

\ead{rafael.zadorosny@unesp.br}

%%%%%%

\begin{abstract}

Although the production of YBa$_{2}$Cu$_{3}$O$_{7-\delta}$ (Y123) has been extensively reported, there is still a lack of information on the ideal heat treatment to produce this material in the form of one dimension nanostructures. Thus, by means of the Solution Blow Spinning technique, metals embedded in polymer fibers were prepared. These polymer composite fibers were fired and then investigated by thermogravimetric analysis. The maximum sintering temperatures of heat treatment were chosen in the interval \SI{850}{\celsius}-\SI{925}{\celsius} for one hour under oxygen flux. SEM images allowed us to determine the wire diameter as approximately 350~nm for all samples, as well as to map the evolution of the entangled wire morphology with the sintering temperature. XRD analysis indicated the presence of Y123 and secondary phases in all samples. Ac magnetic susceptibility and dc magnetization measurements demonstrated that the sample sintered at \SI{925}{\celsius}/1h is the one with the highest weak-link critical temperature and the largest diamagnetic response.

\end{abstract}

\vspace{2pc}

%\submitto{\SUST}

\section{Introduction}
\label{S:1}
%due to their different
The study of materials at the nano- and submicrometric scales has been of great interest as they exhibit different behaviors compared to their respective bulk specimens~\cite{vakifahmetoglu2011fabrication,wang2013situ,Jeevanandam2018nano}.
In particular, superconducting materials of nanometric sizes have been used as elements of electronic circuits~\cite{braginski2019superconductor}, single-photon detectors~\cite{gaudio2014inhomogeneous}, and also quantum batteries~\cite{strambini2020battery}. Despite being one of the first high-temperature superconductors (HTS) discovered, YBa$_{2}$Cu$_{3}$O$_{7-\delta}$ (Y123) has still been the subject of several studies~\cite{antal2020relationship,rijckaert2020superconducting,heyl2020dissolution}. Trabaldo \textit{et. al}~\cite{trabaldo2019transport} have proposed that SQUIDS based on HTS nanowires can be considered potential candidates as highly sensitive magnetometers for low-field magnetic resonance imaging and magnetoencephalography, for instance. 
Another interesting phenomenon predicted for superconducting nanowires is the wireless power transfer which occurs in enhanced HTS capacitors based on double-sided Y123 films~\cite{he2018superconducting}.

Among several techniques to prepare polymeric and ceramic nanowires, Solution Blow Spinning (SBS) is the most simplified and low-cost method~\cite{medeiros2009solution,Song2020,rotta2016ybco,daristole2016rev,rotta2019synthesis}. As compared to other techniques~\cite{duarte2014electrospinning,arpaia2013film}, SBS provides a large-scale production of Y123 wires with average diameters at the submicrometric scale~\cite{rotta2019solution}. In SBS, polymer fibers are produced from a spinnable precursor solution stretched by pressured air towards a collector~\cite{medeiros2009solution}. Since different polymers can be used, this technique is more versatile to achieve the initial fibers~\cite{Dias2020polymer}. So, the resulting entangled fibers are firstly fired to eliminate the carbonic material and obtain preceramic fibers. After controlled post-treatments, the final ceramic wires are reached.

From a practical point of view, the current which HTS granular superconductors can carry is limited by weakly coupled grain boundaries~\cite{Likharev1979,Hilgenkamp2002gb}. This intergranular material $-$ or weak-links (WLs) $-$ transports supercurrent densities much lower than the bulk grains, which are pure and crystalline phase. Thus, the magnetic response is the sum of the intra and intergranular contributions. In superconducting ceramics, the intergranular regions arise from several contributions, like grain boundaries formed by non-reacted or non-superconducting material, secondary or weakened superconducting phases, and other structural defects, which are highly dependent on the material synthesis and processing conditions. Among other processing parameters, the sintering temperature has been recognized as one of the most important steps to improve the intergranular properties in HTS. Concerning heat treatments in Y123 bulk ceramics, several investigations have been done since the late 1980's (see, for instance, Pathak and Mishra~\cite{pathak2005review} and references therein). Despite all this progress, the best heat treatment employed to produce Y123 superconducting nanowires is still an open subject of great technological interest.

Having in mind the optimization of the weak-link response in such superconducting submicron specimens, we investigated systematically different heat treatments to produce Y123 superconducting nanowires. The samples were prepared using the SBS technique and sinterized under different temperatures, from \SI{850}{\celsius} to \SI{925}{\celsius}, keeping the same atmosphere, rate of temperature change, and time spent at the plateau temperature. The fibers were characterized by Thermogravimetry (TG/DTG), Scanning Electron Microscopy (SEM), X-ray diffraction (XRD), as well as dc and ac magnetometry.

\section{Materials and Methods}
\label{S:2}

For the synthesis of the precursor solution, 59.06~wt\% methanol (VETEC), 18.12~wt\% acetic acid (Sigma-Aldrich), and 22.82~wt\% propionic acid (Sigma-Aldrich) were used~\cite{rotta2016ybco}. The acetates of yttrium, barium, and copper, all of them from Sigma-Aldrich, were stoichiometrically added to the aforementioned solvents to prepare 2~g of Y123. The solution was polymerized using polyvinylpyrrolidone (Sigma-Aldrich), PVP ((C$_6$H$_9$NO)$_n$, of 1,300,000~g/mol), in the weight ratio~5:1 [acetates:PVP]. The concentration of PVP in the solutions was 5~wt\%. 

\begin{figure}[ht!]
\centering\includegraphics[width=1\linewidth]{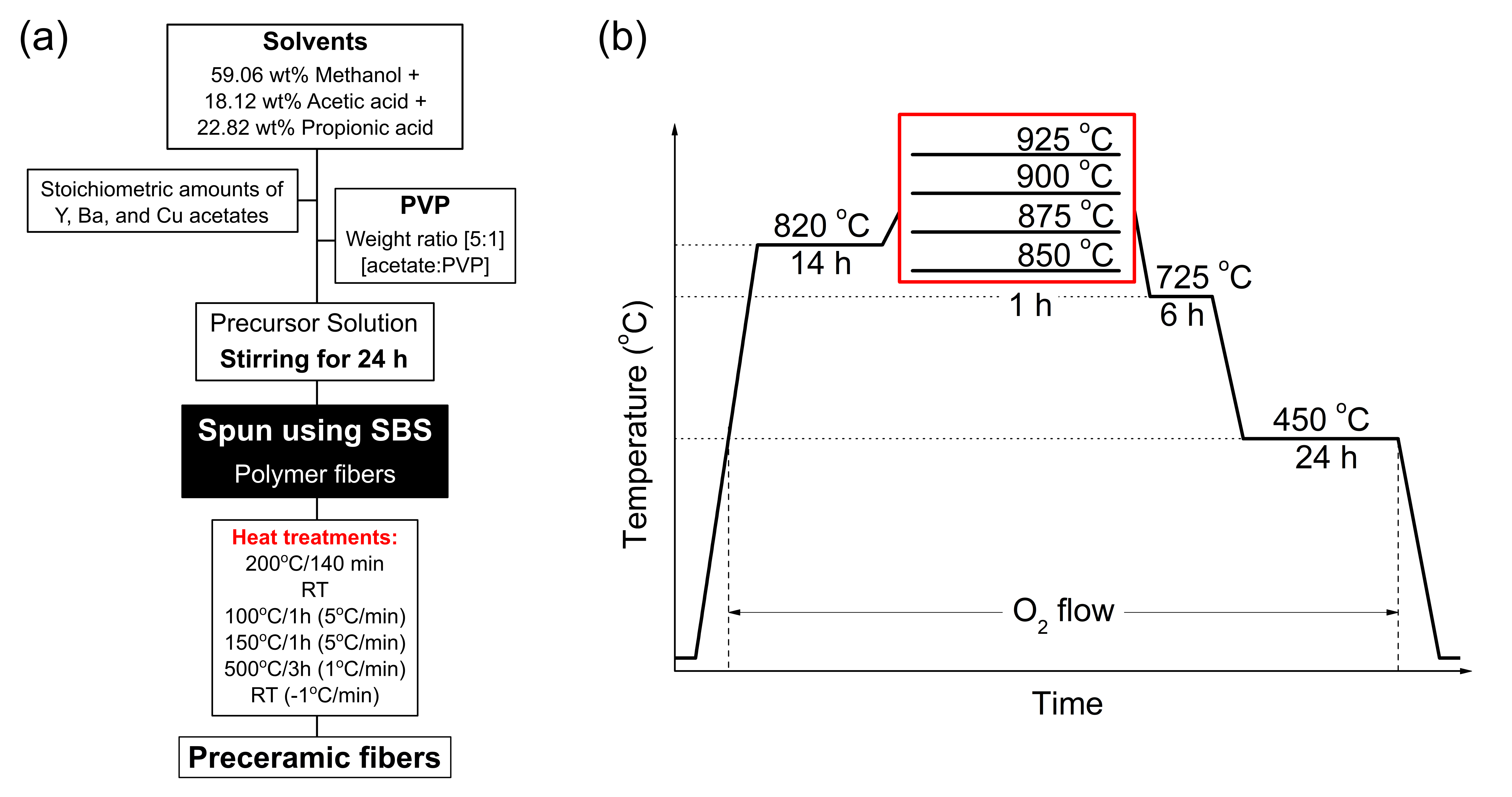}
\caption{Sample preparation details. (a) Flowchart of the preceramic wire preparation. (b) Heat treatment for the samples S850, S875, S900, and S925.}
\label{Fig:HeatTreatments}
\end{figure}

The precursor solution was left under magnetic stirring for 24 hours at room temperature (RT, \SI{25}{\celsius}) in a hermetically sealed vessel. The fibers were spun using the SBS technique~\cite{medeiros2009solution,rotta2016ybco,rotta2019solution,rotta2019synthesis}. The SBS parameters were kept constant, using a 25G-type needle with 0.5~mm external diameter, a cylindrical collector rotating at 40~rpm, a working distance between the needle and the collector of 30 cm, and the solution injection rate of 50~$\mu$l/min. The temperature along the working distance was kept at 30-\SI{35}{\celsius} using a 150~W halogen lamp in this region which aids the evaporation of the solvents. The collected polymer fibers distributed as a non-woven wire network were subjected to a heat treatment in a domestic oven at \SI{200}{\celsius} for 140~min in order to eliminate the remaining methanol, acetic, and propionic acids. In a subsequent heat treatment, the precursor ceramic was obtained by decomposing the organic components as follows: at \SI{5}{\celsius}/min from RT to \SI{100}{\celsius}/1~h; at the same rate up to \SI{150}{\celsius} for another 1~h; and at \SI{1}{\celsius}/min to \SI{500}{\celsius} for 3~h. The cooling rate was \SI{-1}{\celsius}/min to RT. Details of the sample preparation are also presented in Fig.~\ref{Fig:HeatTreatments}(a).

After that, the resulting material was divided in batches and each one was heat treated at the sintering temperatures of \SI{850}{\celsius} (S850), \SI{875}{\celsius} (S875), \SI{900}{\celsius} (S900), and \SI{925}{\celsius} (S925) for 1 hour. Such treatments, also described in Refs.~\cite{rotta2016ybco,rotta2019solution}, were performed in a properly calibrated tubular furnace from room temperature to \SI{820}{\celsius} at \SI{3}{\celsius}/min, followed by a plateau of 14~h; then, heating at \SI{1}{\celsius}/min until each of the sintering temperatures described above. Finally, cooling at the rate \SI{-1}{\celsius}/min up to \SI{725}{\celsius}/6~h, then at \SI{-3}{\celsius}/min to \SI{450}{\celsius}/24~h, and to room temperature at \SI{-3}{\celsius}/min. The $O_2$ flow was switched on at \SI{450}{\celsius} when the temperature was ramping up, and it was switched off after 24~h at \SI{450}{\celsius} during the cooling procedure. Finally, the ceramic wires were obtained. The steps of the heat treatment are schematically shown in Fig.~\ref{Fig:HeatTreatments}(b).

Thermogravimetric measurements were performed using TA Instruments equipment, model Q600, heating 8.354~mg of the non-woven polymeric fibers from room temperature to \SI{1200}{\celsius} at \SI{10}{\celsius}/min in an alumina crucible under nitrogen atmosphere with a flow rate of 100~ml/min. SEM images were obtained in a Zeiss Scanning Electron Microscope, model EVO LS15, operating between 15~kV and 20~kV. In all cases, the specimens were fixed on a sample holder with carbon conductive tape and covered with a gold thin layer to improve their conductivity. In order to identify the crystallographic phases, XRD measurements were performed in a Shimadzu diffractometer, model XRD-6000, with Cu-K$\alpha$ radiation ($\lambda$ = \SI{1.54056}{\angstrom}), a scanning speed of 1$^{\circ}$/min, and measuring in steps of 0.02$^{\circ}$ in the range 15$^{\circ}$ $\le$ 2$\theta$ $\le$ 65$^{\circ}$. 

In order to obtain the superconducting critical temperature (T$_c$) and the weak-link superconducting critical temperature (T$_{c,wl}$), ac susceptibility $\chi_{ac}=\chi '+i\chi ''$ measurements were carried out as a function of temperature at dc remanent field using a Quantum Design MPMS-5S SQUID magnetometer. The remanent field is tipically of the order of 1-2~Oe in the experimental station employed. The ac field amplitude $h$ ranged from 0.05~Oe to 3.8~Oe and the frequency was set to $f =$ 100~Hz. Additionally, temperature-dependent dc magnetization measurements were performed under zero-field-cooling (ZFC) and field-cooled cooling (FCC) conditions at a dc field $H =$ 10~Oe.

\section{Results and discussion}
\label{S:3}

\subsection{Thermal Analysis}

\begin{figure}[h!]
\centering\includegraphics[width=0.6\linewidth]{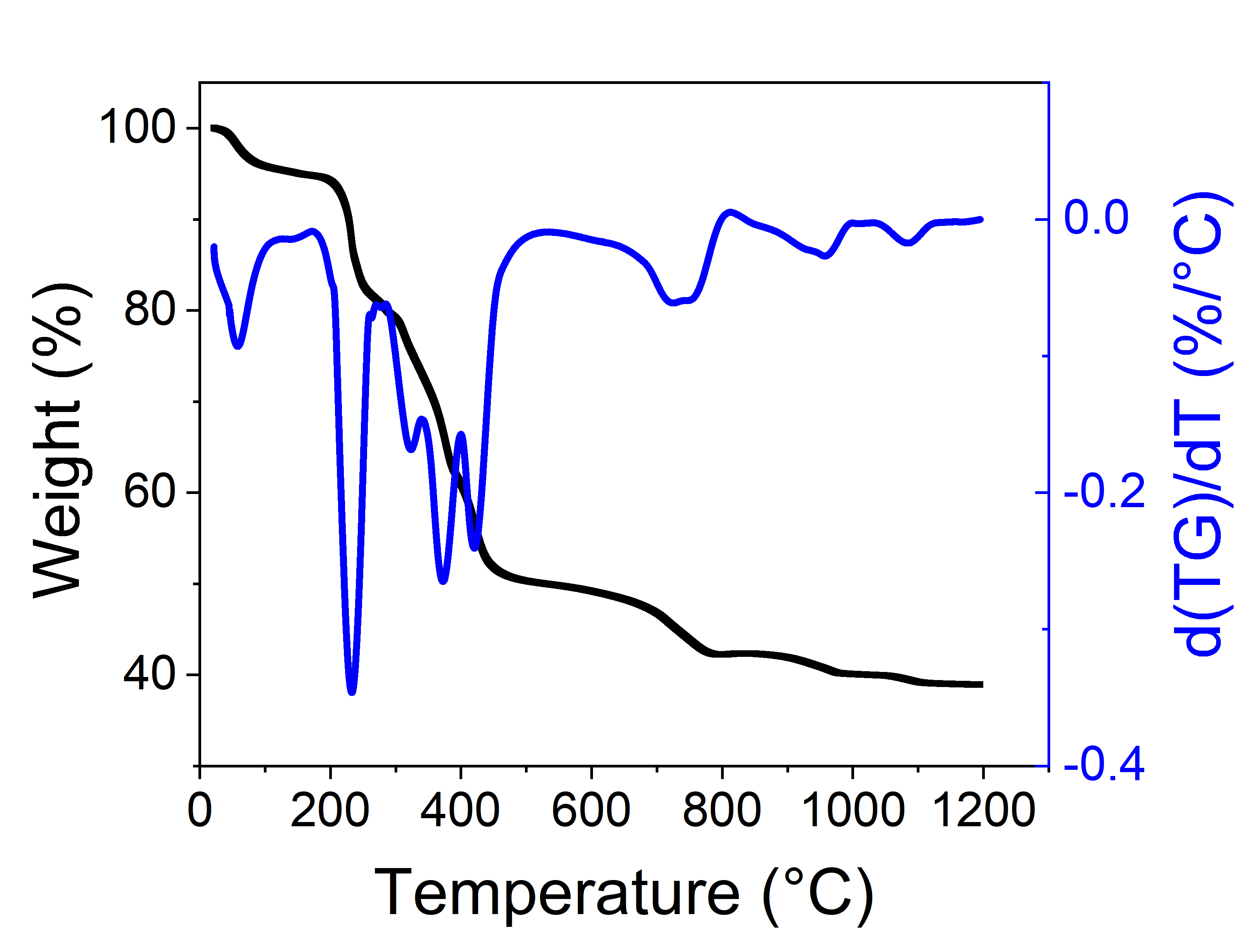}
\caption{Thermogravimetric analysis (TG and DTG curves) of the polymeric precursor sample. Up to \SI{500}{\celsius}, the weight loss and the DTG dips are related to water, solvent evaporation, and polymer elimination. Dips above \SI{800}{\celsius} are related to crystallization of Y123 compounds.}
\label{Fig:TG-DTG}
\end{figure}

Thermal information on the polymeric fibers was obtained by thermogravimetry analysis. Fig.~\ref{Fig:TG-DTG} shows the dependence of weight loss (TG), and its derivative, DTG, on temperature.
When compared to other chemical methods~\cite{motta2008chel}, in which an intermediate polymer is also added and removed afterwards, a larger amount of mass is lost up to \SI{500}{\celsius} ($\approx$ 50\%). From room temperature to approximately \SI{300}{\celsius}, water absorbed by the highly hygroscopic polymer PVP is eliminated, as well as volatile and organic solvents~\cite{duarte2014electrospinning}. The dips in the DTG curve ranging from \SI{300}{\celsius} to \SI{500}{\celsius} refer to the decomposition of organic groups and of the polymer itself~\cite{yuh2007sol-gel,duarte2014electrospinning,rotta2016ybco}. For temperatures above \SI{700}{\celsius}, less-significant weight losses were observed ($\approx$ 10\%), and the corresponding peaks in DTG were due to the formation of intermediate chemical compounds associated with Y123, as well as the Y123 phase itself. Similar behavior was observed by Jasim \textit{et al.}~\cite{jasim2016fabrication} and Pathak and co-workers~\cite{pathak2005review}, who describe the beginning of the YBCO crystallization process at \SI{820}{\celsius} and \SI{850}{\celsius}, respectively. Therefore, we chose \SI{850}{\celsius} as the minimum sintering temperature and \SI{925}{\celsius} as the maximum one, since the Y123 phase starts its decomposition process above this value~\cite{yahya2012y123}. 

It is worth mentioning that the final weight loss was 59\% up to \SI{925}{\celsius}, similar to the values obtained by Duarte \textit{et al.}~\cite{duarte2014electrospinning} and Rotta \textit{et al.}~\cite{rotta2016ybco} of 57\% and 64\%, respectively. Although the molecular weight of PVP is different from Ref.~\cite{rotta2016ybco}, such similarity is expected due to the equal acetate to polymer weight ratio.

\subsection{Scanning Electron Microscopy}
\label{S:32}

\begin{figure}[ht!]
\centering\includegraphics[width=0.8\linewidth]{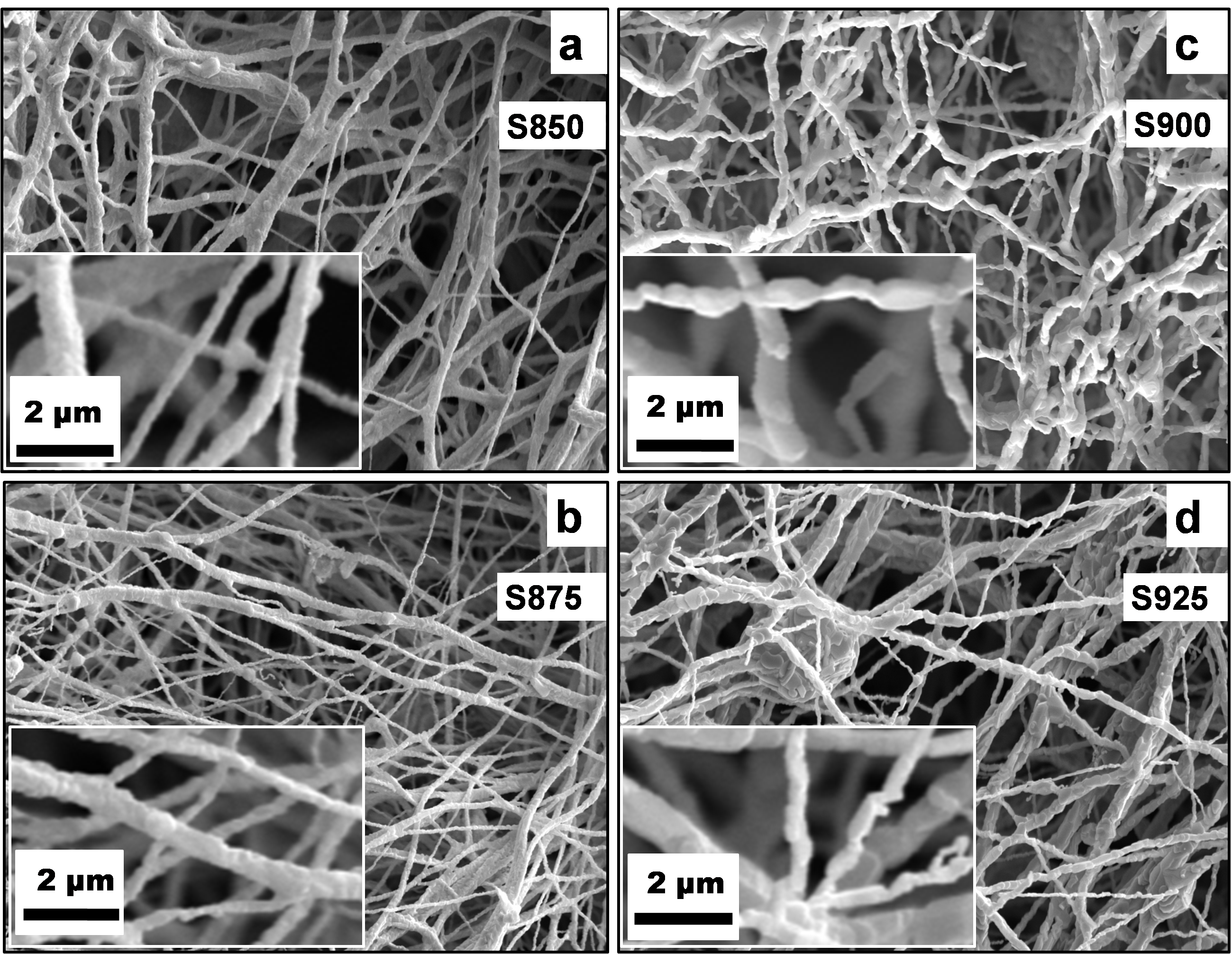}
\caption{SEM images of Y123 samples subjected to different maximum sintering temperatures (a) \SI{850}{\celsius}, (b) \SI{875}{\celsius}, (c) \SI{900}{\celsius} and (d) \SI{925}{\celsius}.}
\label{Fig:MEV}
\end{figure}

\begin{table}[]
\centering
\caption{Representative dimensions obtained from the SEM images. The average dimensions $d_{w}$, $d_{g}$, and $h_{c}$ are shown for each specimen. The $d_{g}$ distribution for S850 and S875 ranges from 131 to 573 nm and from 128 to 758 nm, respectively. The $h_{c}$ distribution for S900 and S925 has a wider range from 214 to 1301 nm and from 233 to 1380 nm, respectively.}
\label{Tab:AverDiam}
\begin{tabular}{c|ccc}
\hline
\textbf{Samples} & \textbf{$d_{w}$ (nm)} & \textbf{$d_{g}$ (nm)} & \textbf{$h_{c}$ (nm)} \\ \hline
S850 & 330 & 267 & -   \\
S875 & 372 & 278 & -   \\
S900 & 376 & -   & 490 \\
S925 & 359 & -   & 533 \\ \hline
\end{tabular}
\end{table}

The morphology of the specimens was investigated by SEM images (Fig.~\ref{Fig:MEV}). The ceramic wires had an average diameter ($d_{w}$) ranging from 330~nm to 376~nm, as shown in Table~\ref{Tab:AverDiam}, with standard deviations around 150~nm. These values were obtained from a Gaussian approach of the diameter distribution by counting a hundred wires taken randomly (see Ref.~\cite{rotta2016ybco}). The large deviations are in par with the wide range of diameters noticed for the wires within the same sample, ranging from 100 nm to 1500 nm. As can be observed, $d_w$ is similar among the produced samples. Therefore, it means that different sintering temperatures do not change $d_w$ significantly, which only depends on the initial parameters of the precursor solution and spinning process, such as, viscosity, injection rate, and pressure~\cite{rotta2019solution}.

Fig.~\ref{Fig:MEV} shows SEM images for the specimens S850 [panel (a)], S875 (b), S900 (c), S925 (d). In all samples, the ceramics are distributed as in a web of entangled wires. They are composed of small grains connected along an axis, resulting in rough surfaces. The insets show smaller grains for the samples S850 and S875 in comparison to those of S900 and S925. Thereby, the wire surface becomes smoother for higher sintering temperatures.

It is well-known that the driving force for the sintering process is the reduction of the interfacial free energy of a granular system~\cite{kingery1976ceramics}. Therefore, grains that are in intimate contact with each other grow along these connections, forming necks among them and finally coalescing ideally as one. In our case, the wires appear to be as a 3D array of grains distributed along the wire axis for the sintering temperatures of \SI{850}{\celsius} and \SI{875}{\celsius}. At these temperatures, grains are nearly rounded and their average diameter ($d_g$) is shown in Table~\ref{Tab:AverDiam}. As expected, $d_g$ increases for higher temperatures, however, its is still smaller than the wire diameter. For the samples S900 and S925, there is a change in the overall grain shape, which becomes elongated along the wire axis, as a consequence of the grain growth promoted at higher temperatures. The diameter of the cylinder-like grain is about the same as the wire diameter for both temperatures. For the sample S900, the average height of the cylinder ($h_{c}$) is about 490~nm, although some round grains can still be found in Fig.~\ref{Fig:MEV}(c). These dimensions are listed in Table~\ref{Tab:AverDiam}. For the sample treated at a higher temperature, a highly directional sintering process takes place since the grains become even more elongated ($h_{c} \approx$ 533~nm) and their boundaries are roughly perpendicular to the main axis of the wire. In some regions, a number of stacking grains is found in the SEM images, representing a step forward in the grain growth sintering process. Therefore, increasing the heat treatment maximum temperature allows for the appearance of larger grains, which are able to grow oriented along the wire, at the expenses of smaller ones. A quite similar evolution of the grain morphology was observed in CoFe$_2$O$_4$ nanowires for sintering temperatures from \SI{500}{\celsius} to \SI{850}{\celsius}~\cite{wang2008CoFeO}. The authors explain the kinetics of grain growth as being due to successive Ostwald ripening processes, which also seems to be the case here for nearly-one-dimensional Y123 superconductors.

Therefore, these changes in the microstructure with the sintering temperature affect the amount of grain boundaries, which decreases with increasing temperatures, a feature accompanied by the strengthening of the junctions among the grains. On the other hand, the interconnections of the wires arranged as a network are similar for all samples. For superconducting materials, there is a high dependence of the capacity to transport superconducting currents with the connectivity among the grains~\cite{benavidez2005sintering,ozkurteffect} and the wires~\cite{Zeng2017mag}. This is further discussed in Section~\ref{mag}.

\subsection{X-ray Diffractometry}

\begin{figure}[h!]
\centering\includegraphics[width=0.7\linewidth]{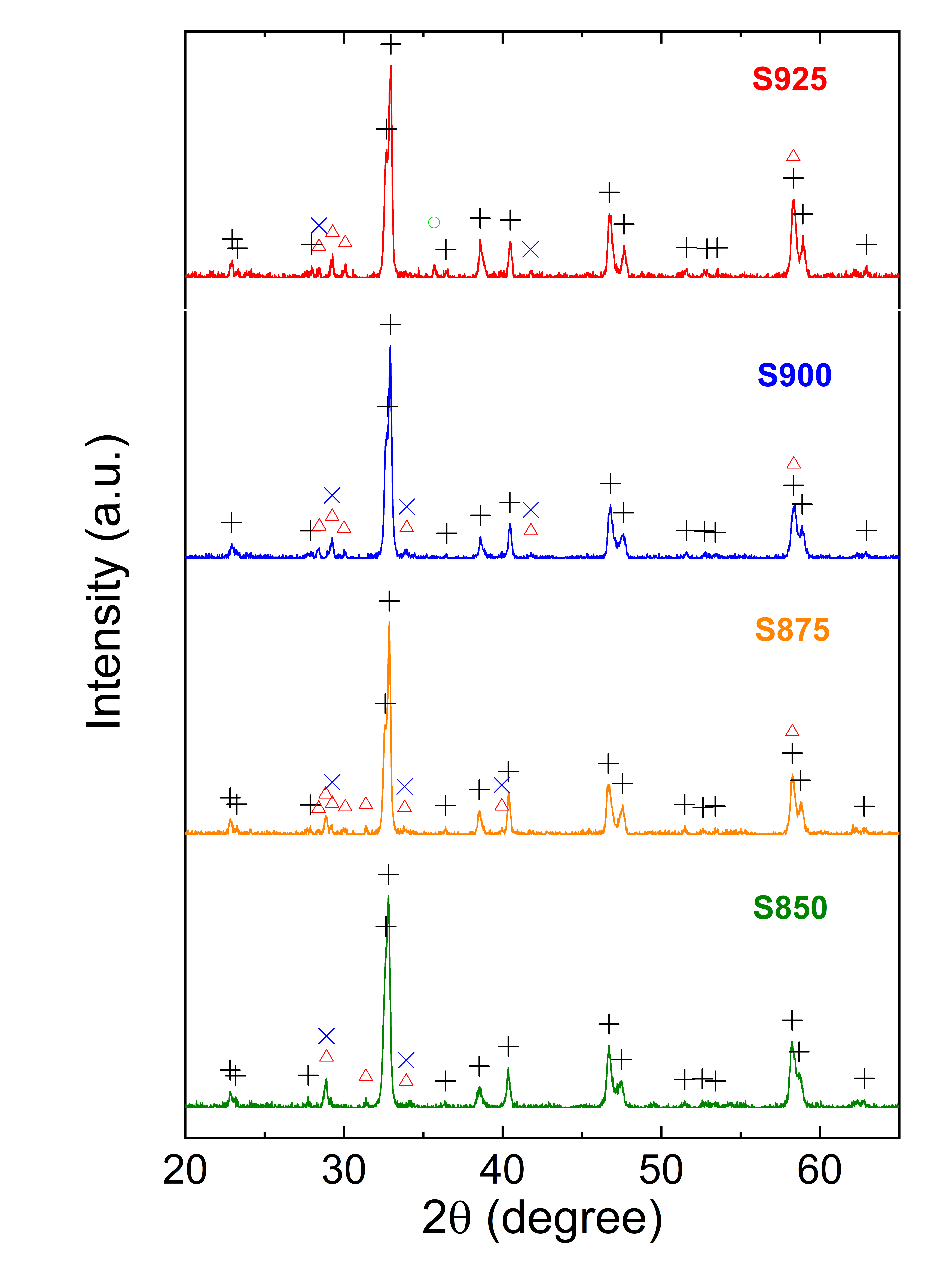}
\caption{XRD analysis. Phase identification for all samples. Symbols represent the folowing phases: Y123 ($+$), BaCuO$_x$ ($\bigtriangleup$), Y$_2$O$_3$ ($\times$), and CuO ($\circ$).}
\label{Fig:Painel_DRX}
\end{figure}

The XRD diagrams shown in Fig.~\ref{Fig:Painel_DRX} clearly demonstrate that the most prominent phase in all samples is the orthorhombic Y123 phase (JCPDS card number 78-2269, space group $Pmmm$) and its peak intensities at different sintering temperatures do not change significantly. Besides that, secondary phases of Y$_2$O$_3$ (JCPDS 71-0099), BaCuO$_x$ (JCPDS 38-1402 for $x$ = 2 and JCPDS 49-0149 for $x$ = 2.5), and CuO (JCPDS 65-2309) were identified. These spurious phases are intermediate compounds which react among themselves to produce YBa$_2$Cu$_3$O$_{7-\delta}$~\cite{agostino2004ba2cu3o7}.

\subsection{Magnetic Characterization}
\label{mag}

\begin{figure}[b!]
\centering\includegraphics[width=1.0\linewidth]{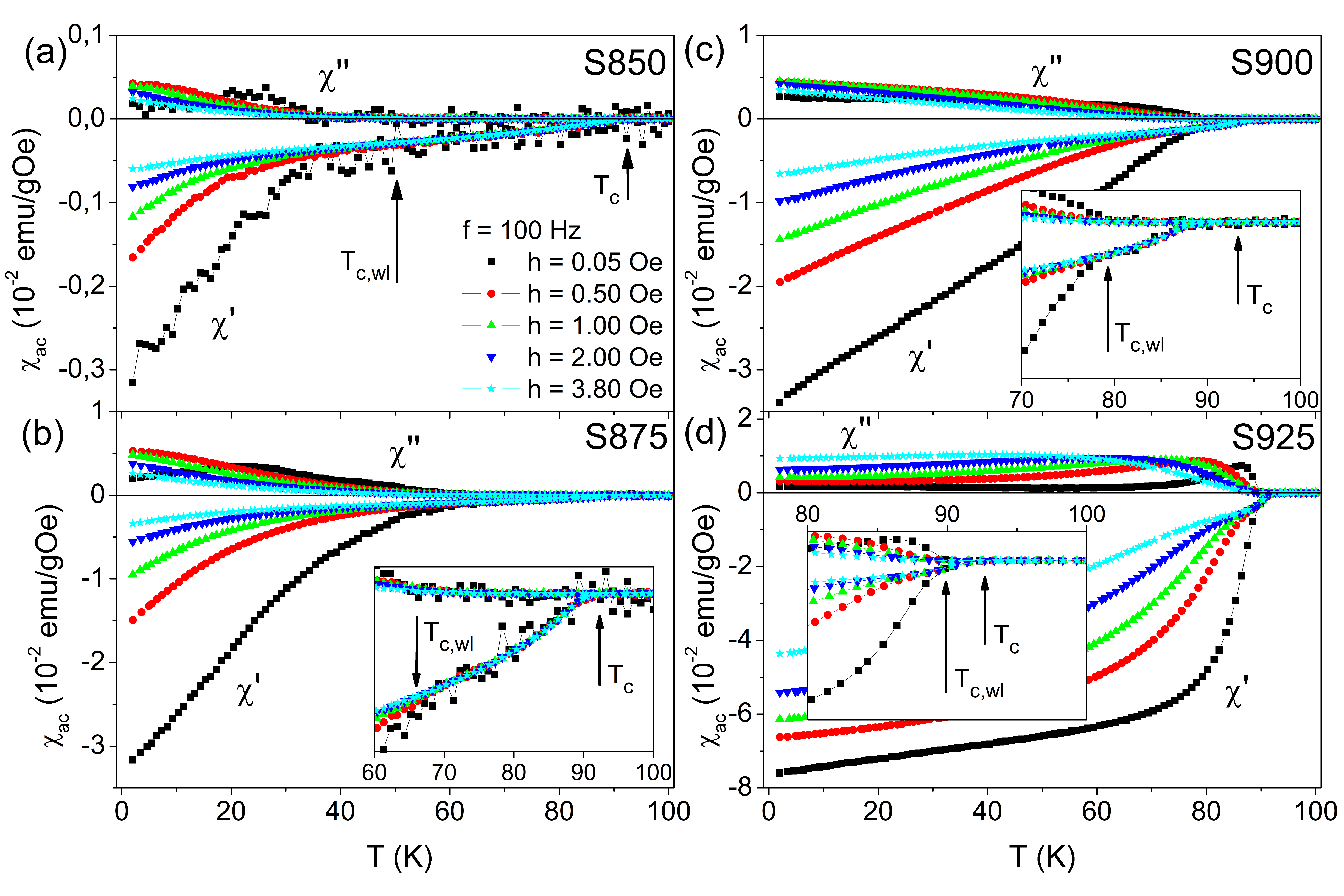} 
\caption{$\chi_{ac}(T)$ measurements for $h=$ 0.05, 0.5, 1, 2, and 3.8 Oe at 100 Hz and remnant dc field for the samples (a) S850, (b) S875, (c) S900, and (d) S925.}
\label{Fig:chi}
\end{figure}

For the investigations of granularity in HTS, ac magnetic susceptibility has proved to be a powerful tool~\cite{gomory1997acsuscept,Passos2001granul}. In this technique, the real ($\chi '$) and imaginary ($\chi ''$) parts are the in-phase and out-of-phase components, respectively. The first one is related to the shielding capacity due to the superconducting screening currents, whereas the second one is associated with ac losses in the superconductor. The mass-normalized $\chi_{ac}$(T) measurements were carried out for some ac fields at 100 Hz and remnant dc field for all ceramic samples, as presented in Fig.~\ref{Fig:chi}. As expected, when the ac-field amplitude $h$ increases, the overall shielding capacity associated with $\chi '$ decreases for each one of the specimens accompanied by an increase in ac losses given by the $\chi ''$ component. It is important to mention that the curves taken at $h=$ 0.05 Oe for the samples S850 (panel (a)) and S875 (inset in (b)) are noisy due to the low signal-to-noise ratio. Moreover, there is an improvement of the maximum shielding by almost 30 times from the S850 to the S925 specimen. 

In order to determine T$_c$, both ac susceptibility components, $\chi '$ and $\chi''$, were examined from higher to lower temperatures. Subsequently, the transition temperature was chosen as the value below which they detached from each other. Similarly, the onset weak-link critical temperature T$_{c,wl}$ was chosen as the value below which $\chi '$ curves become $h$-field dependent~\cite{Godfarb1991magnetic}. Table~\ref{Tab:Tc} summarizes both transition temperatures for all samples. One can note that T$_c$ shows a slight variation within the experimental error around 92~K. It means that the grains are somewhat homogeneous among different sintering temperatures, i.e., the oxygen stoichiometry is essentially equal inside the grains. On the other hand, T$_{c,wl}$ varies drastically and monotonically from 50~K to 90~K from sample S850 to S925. Therefore, the sintering temperature strongly influences the intergrain superconducting response as higher temperatures improve the transport of supercurrents among the grains which compose the wires. Moreover, this result corroborates the fact that increasing the sintering temperature diminishes the density of WLs, as discussed in Section~\ref{S:32}, which also become stronger, resulting in an improvement of the total magnetic response of the sample.

Futhermore, the curve $h=$ 0.05~Oe for the sample S925 (Fig.~\ref{Fig:chi}(d)) shows a sudden decrease in the in-phase component below T$_{c,wl}$ together with a well-defined peak in $\chi ''$, but magnetic flux is not entirely expelled from the intergranular regions. For lower temperatures, some weak-links still switch on progressively. For higher amplitudes, some of those links are so weak that they significantly reduce the superconducting response or simply do not undergo the superconducting transition. This behavior is even more significant for the samples S850, S875, and S900 since the WLs are weaker. In other words, the intergrain goes to the superconducting state progressively, resulting in a broad transition width which extents along the entire temperature range. This feature is common in granular systems, notably when there is no 3D densification of the ceramic specimen~\cite{motta2008chel}.

It is worth mentioning that the secondary phases are essentially the same. Besides that, the wires are entangled and huge voids are formed among the fibers, playing an important role in the magnetic response as they act as places where flux is trapped~\cite{Zeng2017mag}. Nonetheless, their effects are likely to be similar on average for all samples. Therefore, the density of grain boundaries, which act as weak-links, are the main ingredient for the different behavior between these samples.

\begin{table}[t]
\caption{Critical temperatures of the grain itself (T$_c$) and the weak-links (T$_{c,wl}$) obtained from ac magnetic susceptibility for the ceramic samples.}
\label{Tab:Tc}
\centering
\begin{tabular}{c|cc}
\hline
\textbf{Sample} & \textbf{T$_c$ ($\pm$ 0.5~K)} & \textbf{T$_{c,wl}$ ($\pm$ 0.5~K)} \\ \hline
S850   & 92.5                & 50.3                     \\
S875   & 92.3                & 66.0                     \\
S900   & 93.3                & 79.3                     \\
S925   & 92.7                & 89.9                     \\ \hline
\end{tabular}
\end{table}

\begin{figure}[b!]
\centering\includegraphics[width=1\linewidth]{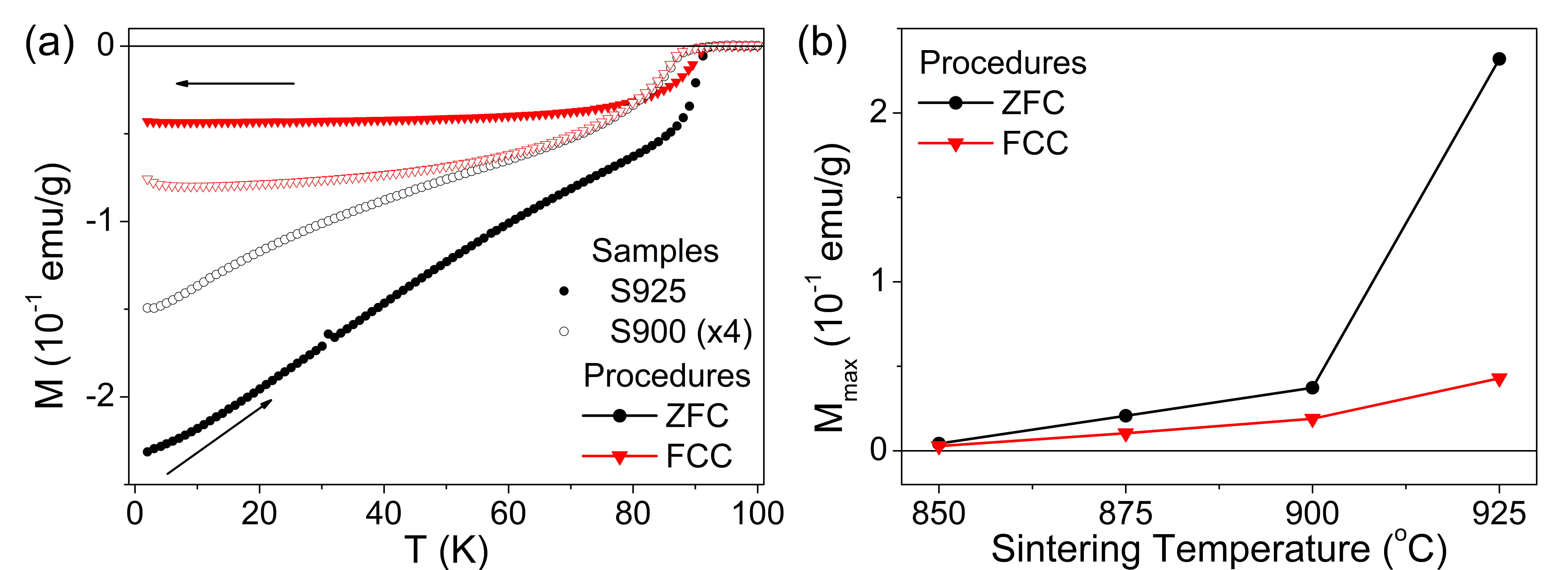}
\caption{Dc magnetization response. (a) Temperature-dependent magnetization curves for the samples S900 ($\times$ 4 for better visualization) and S925 taken following ZFC and FCC procedures. (b) Absolute value of the dc magnetization at $T=$ 2~K obtained from ZFC and FCC curves as a function of the sintering temperature. In these cases, magnetization is given by the ratio of the magnetic moment to the sample mass.}
\label{Fig:M(T)}
\end{figure}

Fig.~\ref{Fig:M(T)}(a) shows dc-magnetization (M) versus temperature curves taken following the ZFC and FCC procedures~\cite{Clem1993zfcfcctrm} for the samples S900 and S925. For better visualization, the data for specimen S900 were multiplied by a factor 4. As long as a dc field is applied ($H=$ 10 Oe), after the sample reaches the lowest temperature in the ZFC curves, regions of the samples are shielded and there is a reduction in the shielding currents as the temperature increases until T$_c$. Subsequently, the FCC procedure is performed by keeping the same applied field above T$_c$ and then the magnetization is taken by cooling the sample down. In this case, magnetic flux is partially expelled from the grains due to the shielding currents, but some remains frozen in non-superconducting or weakened superconducting regions surrounded by or inside the grains. It occurs because the gradient of $H$ is not large enough for depinning the flux. As a consequence, magnetization shows a plateau at low temperatures~\cite{Clem1993zfcfcctrm}. The overall behavior of those curves is identical for the samples S850 and S875 (not shown).

In order to compare all ceramic wires, the absolute value of the magnetization, M$_{max}$, at the lowest temperature (2~K) for each procedure (ZFC and FCC) was chosen as representative values and are plotted against the sintering temperature in Fig.~\ref{Fig:M(T)}(b). As can be seen, M$_{max}$ increases with the sintering temperature for all procedures. There is, however, a substantial increase on the ZFC response of sample S925, which is likely due to an improvement in the grain and intergrain responses, reinforced by a decrease in the density of WLs. Therefore, the screening currents shield the sample more efficiently. On the other hand, the values of M$_{max}$ change slightly for the FCC condition. It means that the increase of the shielding currents, promoted by the higher sintering temperatures, is not completely matched by the lower ability of the samples to pin and trap flux. Although the density of WLs decreases for sample S925, the overall response is composed by several other contributions, such as the defects in the grains, as well as the huge void structure with several interconnections among the individual wires.

Furthermore, the maximum shielding behavior is similar for both magnetic techniques, i.e., dc- and ac-magnetometry. The difference lies on the fact that ac fields are more damaging to the magnetic response of superconductors, having greater penetration and dynamic effects. WLs are then weakened by increasing the magnetic field intensity, deteriorating the superconducting properties. Therefore, the trend observed in Fig.~\ref{Fig:chi} reinforces the idea that increasing sintering temperature improves the WLs. Thus, the results obtained from the magnetic measurements corroborate what is observed in the SEM images, i.e., rod grains forming smooth wire structures can be associated with a better sintering process, improving the grain boundaries and promoting a homogenization and enlargement of the grains. As the maximum superconducting current density J$_c$ that a superconductor can bear is limited by the weak-links in granular superconductors, one can conclude that S925 presents the highest J$_c$ in comparison with the other samples.

\section{Conclusion}
\label{S:4}

The influence of the sintering and heat treatment temperatures on the structural and superconducting properties of Y123 submicrowires was investigated. The samples were prepared using the Solution Blow Spinning technique and, after intermediate heat treatments, sintered at temperatures of \SI{850}{\celsius}, \SI{875}{\celsius}, \SI{900}{\celsius}, and \SI{925}{\celsius} for one hour under oxygen atmosphere. The average wire diameter of the batches studied is around 350~nm for all samples, nearly independent on the sintering temperature. This is not surprising, since the solution and spun parameters are exactly the same for all samples. The SEM images reveal that the grains are roughly rounded and smaller for the specimens S850 and S875. On the other hand, grains have cylinder-like shape for the samples S900 and S925, as a result of the coalescence of two or more smaller grains into a larger piece. These grains connected along one direction constitute the building blocks for the ceramic wires. As a consequence, the grain boundary density is also reduced for higher maximum sintering temperatures. The magnetic response of all batches of nanowires shows essentially the same intragrain superconducting critical temperature T$_c$, however a significant improvement of the weak-link critical temperature T$_{c,wl}$ was observed upon increase of the sintering temperature. Consequently, the superconducting critical currents increase considerably due to stronger links and the sample S925 presented the best results.

\section{Acknowledgements} Brazilian agencies S\~ao Paulo Research Foundation (FAPESP, grant 2016/12390-6), Coordena\c{c}\~ao de Aperfei\c{c}oamento de Pessoal de N\'ivel Superior - Brasil (CAPES) - Finance Code 001, and National Council of Scientific and Technologic Development (CNPq, grants~302564/2018-7, and 130831/2018-2).

AMC: orcid= 0000-0002-4776-5313, credit: Investigation, Data curation, Writing - Original Draft;

DADC: orcid= 0000-0001-6998-3154, credit: Investigation, Data curation, Writing - Review and Editing, Validation;

ALP: orcid= 0000-0002-7949-4626, credit: Investigation, Data curation;

CLP: credit: Data curation, Writing - Review And Editing, Validation;

WAO: orcid= 0000-0002-7778-0979, credit: Writing - Review and Editing, Supervision, Validation;

RZ: orcid= 0000-0002-2419-2049, credit: Conceptualization of this study, Methodology, Resources, Writing - Original Draft, Supervision, Project administration, Funding acquisition;

MM: orcid= 0000-0002-5494-7705, credit: Conceptualization of this study, Methodology, Writing - Original Draft, Writing - Review and Editing, Supervision, Validation.

\section*{References}

\bibliographystyle{iopart-num}
\bibliography{bibliography.bib}

\end{document}